\documentclass[12pt]{article}
\usepackage{epsfig}
\newcommand{\be}{\begin{equation}}
\newcommand{\ee}{\end{equation}}
\newcommand{\ba}{\begin{eqnarray}}
\newcommand{\ea}{\end{eqnarray}}

\begin{document}
\title{Measurements of Protein-Protein Interactions by
Size Exclusion Chromatography}
\author{ J.Bloustine, V.Berejnov\thanks{Present Address: Physics Department, Cornell University, Ithaca, NY 14853},
 S.Fraden \\
Complex Fluids Group, \\The Martin J. Fisher School of Physics,\\
Brandeis University, Waltham, MA 02454} \maketitle

\abstract{A method is presented for determining second virial
coefficients ($B_2$) of protein solutions  from retention time
measurements in size exclusion chromatography (SEC). We determine
$B_2$ by analyzing the concentration dependance of the
chromatographic partition coefficient. We show the ability of this
method to track the evolution of $B_2$ from positive to negative
values in lysozyme and bovine serum albumin solutions. Our SEC
results agree quantitatively with data obtained by light
scattering.}

\section*{Introduction}

It is well known in size exclusion liquid chromatography (SEC)
that the solute retention time depends sensitively on the solute's
size, although no universal calibration for SEC has yet been
achieved. It has also been realized that thermodynamic
non-ideality leads to \emph{concentration} dependent retention
times (Nichol et al.,~1978). Such dependence can be utilized to
quantify the second osmotic virial coefficient, $B_2$.

For a non-ideal solution the osmotic pressure $\Pi$ can be written
as a power series expansion in the solute number density $\rho$
(Hill, 1960).
\begin{equation}                                      \label{virial1}
 \frac{\Pi}{k_{B}T}=\rho+B_2(T)\rho^2+\ldots
\end{equation}
 In
Eq.\ref{virial1} $T$ is the absolute temperature and $k_B$ is
Boltzmann's constant. All terms higher than first order in density
represent non-ideality.

The second virial coefficients of protein solutions have generated
a great deal of interest 
since (George and Wilson, 1994) showed a correlation between
protein crystallisability and $B_2$. Their work demonstrated that
many proteins crystallize in conditions where the second osmotic
virial coefficient becomes slightly negative, indicating net
attractive interactions between protein molecules. The most
prevalent experimental procedure for measuring $B_2$ is light
scattering. Additionally, sedimentation equilibrium (Behlke and
Ristau, 1999), osmometry (Moon at al., 2000), neutron (Velev et
al., 1998) and x-ray scattering (Bonnet\'{e} et al., 1999), and
self-interaction chromatography (Tessier et al., 2002) have been
employed to quantify protein solution non-ideality.

(Nichol et al., 1978) showed the possibility of measuring $B_2$
with frontal elution liquid chromatography. Although frontal
chromatography (Nichol et al.,1978, Wills et al., 1980) allows one
to fix the solute concentration in the column directly, it
requires a large amount of protein ($\sim0.5 \,g$) and long
experiment times (about three hours per column run). In this study
we extend their method to pulse size exclusion HPLC, where a small
amount of protein is injected into and subsequently flows down the
column. This adaptation drastically reduces the amount of protein
($<25 \,$mg) and time needed (about 15 minutes per column run) to
measure $B_2$ by SEC. We show that our results for $B_2$ obtained
with size exclusion chromatography agree well with those from
frontal chromatography and from light scattering measurements. We
also demonstrate that SEC can track the evolution of $B_2$ from
positive to negative values.

\section*{Theory}

For the reader's convenience we reproduce the theory of (Nichol et
al.,1978). We assume a balance of the solute, i.e. protein,
chemical potentials ($\mu_p$ and $\mu_i$) between the stationary
and mobile phases as the solute is transported through the column.
The pore volume (i.e. stationary phase) is labelled with the
subscript $p$, and the inter-pore volume (i.e. mobile phase) with
the subscript $i$. Equilibrium requires $\mu_p=\mu_i $. We write
these chemical potentials by including the standard part $\mu^o$,
the ideal term, and a term accounting for thermodynamic
non-ideality through the activity coefficient $\gamma$:
\begin{eqnarray*}
\mu_p&=&\mu^o_p+RT \; \ln (C_p \,\, \gamma_p(C_p)) \\
\mu_i&=&\mu^o_i+RT \; \ln (C_i \,\, \gamma_i(C_i))
\end{eqnarray*}
where $C_{i,p}$ are the local solute weight concentrations, $R$ is the universal gas constant and
$\gamma_p(C_p), \; \gamma_i(C_i)$ are the thermodynamic activity
coefficients of solute molecules in the pore and inter-pore
volumes respectively. Rearrangement of these equations yields:
\begin{eqnarray}        \label{eq:gamma-i-p}
\ln (K_{0})=\frac{\mu^{o}_{i}-\mu^{o}_{p}}{RT} \nonumber \\
\ln\left(\frac{\gamma_i}{\gamma_p}\right)=\ln
\left(\frac{C_p}{C_i}\right)-\ln (K_0)
\end{eqnarray}
where $K_0$ is the partition coefficient of solute molecules
between chromatographic phases in the limit of infinite dilution.
The relation between weight concentration, $C$, and number
density, $\rho$, is $\rho=C\frac{N_A}{M_w}$. $N_A$ is Avogadro's
number and $M_w$ is the solute molecular mass. (Nichol et al.,
1978) made a virial expansion of the activity coefficients
\begin{eqnarray}            \label{eq:gamma-virial}
\ln\gamma(z)=2 B_2 (N_A/M_w) C + \footnotesize{\mbox{higher
terms}}
\end{eqnarray}
 We note that this consideration
assumes no difference in the solute-solute interactions in the
mobile and stationary phases.  The local solute distribution
coefficient is $K_D \equiv \frac{C_p}{C_i}$. If $K_D$ is
independent of concentration, as is the case for pulse
chromatography with $B_2 = 0$, or if the concentration is constant
as in frontal elution chromatography, then (Nichol et al., 1978,
Yau et al., 1979)
\begin{equation}                                             \label{eq:Kd}
K_D \equiv \frac{C_p}{C_i}=\frac{t_r-t_o}{t_T-t_o}=\frac{V_r-V_o}{V_T-V_o}
\end{equation}
where $t_r$ and $V_r$ are the solute retention time and volume,
$t_0$ and $V_0$ are the retention time and volume of completely
excluded molecules (i.e. the ``dead'' volume), and $t_T$ and $V_T$
the retention time and volume of completely included molecules
(i.e. the ``total'' volume).  Inserting the definition of
$K_D$(Eq.~\ref{eq:Kd}) and Eq.~\ref{eq:gamma-virial} into
Eq.~\ref{eq:gamma-i-p} and keeping only the first order terms in
concentration one obtains a relation between $K_D$, $B_{2}$, and
$C_i$ valid for frontal chromatography where the concentration
$C_i$ is the plateau value of the solute concentration in the
mobile phase:
\begin{equation}                                             \label{eq:b2sec}
\ln \left(\frac{K_D}{K_0}\right)=2 B_2 \frac{N_A}{M_w}
C_i(1-K_D)
\end{equation}
To adapt this to pulse chromatography we replace the plateau value with the average
concentration $<C_i>$ of the mobile phase in the pulse:
\begin{equation}                                             \label{eq:b2}
\ln \left(\frac{K_D}{K_0}\right)=2 B_2 \frac{N_A}{M_w}
<C_i>(1-K_D)
\end{equation}

Since $<C_i>$ is not directly accessible in a HPLC experiment one
must relate it to measurable parameters. One determines the mass
of solute molecules in the pulse, or migration zone,
($m_{\mbox{\scriptsize zone}}$) by integrating the concentration
as a function of time over the zone volume, i.e. the peak($V_z$).
For our columns, in which there is no irreversible binding of
protein molecules to the column, all the injected molecules are
accounted for by integrating the peak. Therefore the total
injected mass is the same as the total mass in the zone,
$m_{\mbox{\scriptsize inj}}=C_{\mbox{\scriptsize
inj}}V_{\mbox{\scriptsize inj}}=m_{\mbox{\scriptsize zone}}$, but
the concentration of solute in the migration zone is much lower
than the injected concentration because the pulse spreads as it is
transported through the column. The condition for the conservation
of mass of solute molecules in the migration zone (subscript $z$)
is
\begin{eqnarray} \label{eq:masscons}
  m_{i}+m_{p}=m_{\mbox{\scriptsize inj}}\\
  <C_{i}> V_{i} + <C_{p}> V_{p}=C_{\mbox{\scriptsize inj}}V_{\mbox{\scriptsize inj}}=m_{\mbox{\scriptsize inj}}\nonumber
\end{eqnarray}

\begin{figure}[ht]
\centering{ \psfig{file=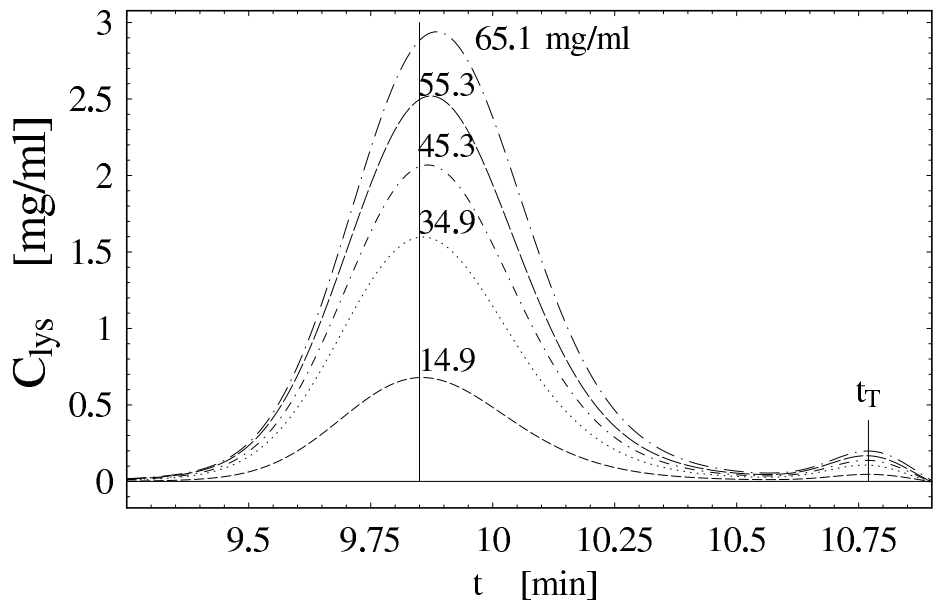,width=15cm}}
 \caption{\,\,Lysozyme chromatograms for
$V_{\mbox{\scriptsize inj}}=20\mu$l and different injected
concentrations ($C_{\mbox{\scriptsize inj}}$) as indicated next to
each curve. The average concentrations $<C_i>$ used in the
analysis in Fig.~\ref{fig:winzor+Kd-vs-Ci} are dash - 0.74~mg/ml,
points - 1.72~mg/ml, dash and points - 2.20~mg/ml, long dash -
2.68~mg/ml, long dash and points - 3.08~mg/ml. The vertical line
marks the retention time  for the most dilute sample (not shown).
The retention time $t_r$ is the time corresponding to the peak of
the concentration profile ($C_{\mbox{\scriptsize max}}$) and
increases with increasing concentration. The retention time of
completely included molecules (the ``total'' volume) is marked as
$t_T$ and is caused by the buffer. The retention time of completed
excluded molecules (the ``dead'' volume) was $t_0 = 6.07$~min (not
shown in figure). Note that $C_{\mbox{\scriptsize max}}$ and
$<C_i>$ are similar. The buffer is Sodium Acetate, 50mM, pH 4.7.
}\label{fig:rawdata}
\end{figure}

Here $V_i$ and $V_p$ are the mobile (inter-pore) and stationary
(pore) portions of the zone volume $V_z$, with
\begin{equation}            \label{eq:Zone-volumes}
\begin{array}{cc}
V_p=(V_T-V_0)\frac{V_z}{V_T} &, \,V_i=(V_0)\frac{V_z}{V_T}
\end{array}
\end{equation}
 We measure the solute zone volume $V_z$ from the full width $\Delta t$ at half-maximum
 of the chromatogram peak using $V_z=\nu\Delta t$, where $\nu$
is the average flow rate. After substituting the definition of the
partition coefficient given in Eq.~\ref{eq:Kd} and definitions
Eq.~\ref{eq:Zone-volumes} into Eq.~\ref{eq:masscons}, one obtains:
\begin{equation}                                             \label{eq:finalconc}
<C_i>=\frac{m_{\mbox{\scriptsize inj}}}{V_z(\frac{V_R}{V_T})}
\end{equation}
A simple way to understand Eq.~\ref{eq:finalconc} is to note that
the numerator is the total mass in the zone and the denominator is
the volume of the zone accessible to the protein. Thus, the
concentration $<C_i>$ is the ratio of these terms. In this
derivation we have assumed Eq.~\ref{eq:Kd} holds, which is no
longer the case when both $B_2 \neq 0$ and the concentration is
changing during transport down the column. However, as we will
show below, the changes in $K_D$ with concentration are small,
which may justify our approximation. This relation allows us to
extend the method of (Nichol et al., 1978), originally developed
using frontal elution chromatography, to pulse HPLC.
Alternatively, one could use the maximum concentration
$C_{\mbox{\scriptsize max}}$ of eluted solute instead of $<C_i>$
in Eq.~\ref{eq:b2}. As shown in Fig.~\ref{fig:rawdata}
$C_{\mbox{\scriptsize max}}$ and $<C_i>$ are almost equal. Our
procedure is then to inject different volumes of samples at
various concentrations, measure $K_D$ from the retention times as
given in Eq.~\ref{eq:Kd} and then plot $\ln K_D$ as a function of
either $<C_i> (1-K_D)$ or $C_{\mbox{\scriptsize max}} (1-K_D)$.
The slope of that plot is then $2 \, B_2 \, N_A/M_w$.

\section*{Experimental}
\subsection*{Materials}
We obtained lysozyme (6x crystallized hen egg white), from
Seikagaku America. Our studies, along with others' (Muschol and
Rosenberger, 1997), of the purity of lysozyme preparations from
Sigma and Seikagaku showed the Seikagaku to be purer and it was
used without further purification. We obtained bovine serum
albumin (BSA), from Sigma, and it was used without further
purification. All buffer components were obtained from Fisher
Scientific. A Millipore Elix system purified water for all the
experiments. We prepared potassium phosphate buffers by mixing
50mM solutions of $\mbox{K}_{2}\mbox{HPO}_{4}$ and
$\mbox{KH}_{2}\mbox{PO}_{4}$, at various NaCl concentrations to
adjust the ionic strength, to reach the desired pH = 6.2 as
measured by an Orion SA520 pH meter. The pH = 4.7 of sodium
acetate buffers was adjusted by adding concentrated acetic acid to
solutions of sodium acetate and NaCl. Additionally all buffers
were passed through 0.45~$\mu$m nylon filters, also obtained from
Millipore, prior to use. Protein concentrations were measured
using a Varian instruments Cary 50Bio spectrophotometer at a
wavelength of 278 nm. The extinction coefficient used for lysozyme
was $\epsilon_{278{\mbox{\scriptsize nm}}} = 2.64$~ml
(mg\,cm)$^{-1}$, and $\epsilon_{278{\mbox{\scriptsize nm}}} =
.667$~ml (mg cm)$^{-1}$ for BSA.

\subsection*{Chromatography}
An 1100 series liquid chromatography (HPLC) system from Agilent
Technologies (Wilmington, DE) was used for all chromatographic
measurements.  Protein retention times were determined using an
Agilent differential refractive index detector (RID) and an
Agilent diode-array-detector (DAD) by absorbance at 278nm. A
TSK-G2000SW (30cm x 0.75cm I.D.) column from TosoBiosep and a
YMC-Diol-200AMP (30cm x 0.60cm I.D.) column from YMC were used in
the chromatographic measurements. We used a flow rate of 1~ml/min
for all measurements.  These columns contain a packing of porous
silica beads whose surfaces have been hydrophilicly modified. From
the manufacturer's specifications the diameter of a single bead is
about $5 \mu$m for both columns. The average pore diameter is 125
\AA~ for the TSK-G2000SW, and 200 \AA~ for the YMC-Diol-200AMP. We
determined the SEC calibration curve for these columns by using
poly-ethylene-glycol (PEG) samples with molecular weights $200
\leq M_w \leq 10^5$ g/mol, obtained from Sigma and Fluka. For
every run the eluent was the same as the sample buffer. The random
run-to-run difference in retention times for our system was
$<0.1\%$. Any dependence of the dimensionless distribution
coefficient $K_D$ for protein molecules between the stationary and
mobile phases on the average flow rate $\nu$ would indicate
non-equilibrium effects. We found $K_D$ to be totally independent
of flow rate for the experimentally accessible values: $0.1 \,
\mbox{ml/min} \, \leq \, \nu \, \leq \, 1.3 \, \mbox{ml/min}$.

\subsection*{Methods}
For each solvent condition we performed a series of HPLC
experiments varying solute (protein) injected concentration
$C_{\mbox{\scriptsize inj}}$ and using two injection volumes,
$V_{\mbox{\scriptsize inj}}=20$ and $100\, \mu$l. We identify the
protein retention time $t_{r}$ as the time of the maximum in the
RID signal (Fig.~1), where the injection time is $t=0$. We plot
$t_{r}$ as a function of $C_{\mbox{\scriptsize inj}}$, and find
that $t_{r}$ depends on $V_{\mbox{\scriptsize inj}}$ as shown in
Fig.~\ref{fig:tr-vs-Cinj}. In order to apply our modification of
(Nichol et al., 1978)'s method to HPLC, we recalculate the average
solute concentration in the peak zone, $<C_i>$, as described in
Eq.~\ref{eq:finalconc}, and find that this reassuringly collapses
the multiple $t_r$ vs. $C_{\mbox{\scriptsize inj}}$ curves from
Fig.~\ref{fig:tr-vs-Cinj} to a single curve as shown in the insert
of Fig.~\ref{fig:winzor+Kd-vs-Ci}. The slope of this collapsed
curve is proportional to the second virial coefficient according
to Eq.~\ref{eq:b2sec}.

In order to calculate $K_D$ according to Eq.~\ref{eq:Kd} we must
measure the total ($t_T$) and dead ($t_0$) times. We have measured
the total time for each run using the solvent peak (these are
maximums of the second peaks ($t_T$) in Fig.~\ref{fig:rawdata}).
In order to measure the dead time, we used PEG with a molecular
weight of $10^5$~g/mol, which is totally excluded from the TSK and
the YMC columns. We have measured the dead times for all solvent
conditions and injection volumes. It is important to measure $t_T$
and $t_0$ separately for all injection volumes to avoid any
instrumental errors associated with precisely identifying the
injection time.

We have performed light scattering measurements to determine
$B_{2}$ independently for a condition where results were not found
in the literature. We employed the same method as in (George and
Wilson, 1994) to measure the Rayleigh ratio of protein solutions
using toluene as a standard at a scattering angle of 90 degrees.
In Eq.~\ref{virial1} $B_2$ has the units of volume, but virial
coefficients are often reported in units of ml mol/g$^2$, which is
denoted by $A_2$ (George and Wilson, 1994). Then
$B_2=A_2M^2_w/N_A$, where $N_A$ is Avogadro's number. Our results
are shown in Table~\ref{tab:SEC+LS}.

\section*{Results}
We have measured the dependance of the retention factor $K_D$ on
$C_{\mbox{\scriptsize inj}}$ and $V_{\mbox{\scriptsize inj}}$ for
lysozyme and BSA in the above mentioned buffers and columns. These
buffer conditions were chosen to investigate the cross-over from
positive to negative $B_2$ values and to compare with data
available in the literature.

Fig.~\ref{fig:rawdata} shows the RID signal measuring the
concentration of the eluted protein versus time for representative
lysozyme chromatograms with $V_{\mbox{\scriptsize inj}}= \,
20\,\mu$l. One can see the retention time increase with increasing
protein concentration, while $t_T$ remains constant.

In the size exclusion mode, the direction of the shift in the
retention time with concentration depends on the sign of $B_2$.
For conditions where $B_2>0$, $t_r$ increases with increasing
protein concentration and where $B_2<0$, $t_r$ decreases with
increasing concentration. If $B_2=0$, $t_r$ is independent of
concentration. Previous studies (Velev et al., 1998, Muschol and
Rosenberger, 1995, Gripon et al., 1997, Kulkarni, 1999) have shown
that $B_2$ for protein solutions depends on the ionic strength of
the solution.

\begin{figure}[ht]
\centering{\psfig{file=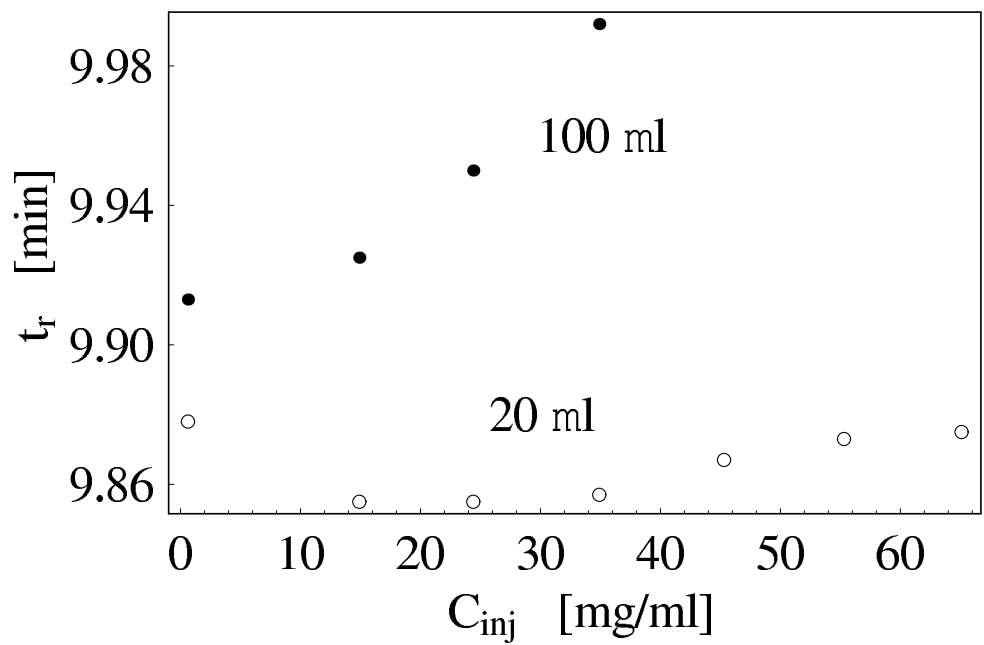,width=15cm}}
 \caption{Lysozyme retention times versus injected
concentrations for two injection volumes. Set $a$~:
$V_{\mbox{\scriptsize inj}}=20 \mu$l and set $b$~:
$V_{\mbox{\scriptsize inj}}=100 \mu$l. Buffer~: Sodium Acetate
50~mM pH 4.7.}\label{fig:tr-vs-Cinj}
\end{figure}

\begin{figure}[ht]
\centering{\psfig{file=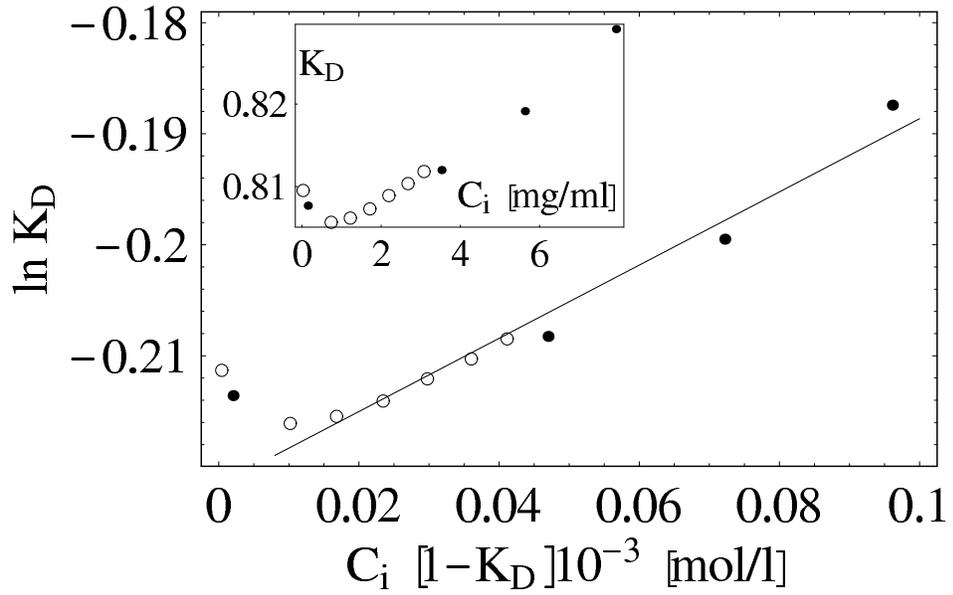,width=15cm}}
 \caption{$\ln K_D$ vs. $<C_i>(1-K_D)$(mg/ml) for
lysozyme as in Eq.~\ref{eq:b2sec}. The insert is a plot of $K_D$
vs. $<C_i>$(mg/ml), where multiple curves from
Fig.~\ref{fig:tr-vs-Cinj} with different injected volumes collapse
after recalculating the solute concentration in the mobile phase
of the migration zone as in Eq.~\ref{eq:finalconc}. Buffer: Sodium
Acetate 50~mM pH 4.7, black dots: $V_{\mbox{\scriptsize inj}} =
100 \mu$l, open circles: $V_{\mbox{\scriptsize inj}} = 20 \mu$l.
}\label{fig:winzor+Kd-vs-Ci}
\end{figure}

Fig.~\ref{fig:tr-vs-Cinj} shows the dependence of lysozyme
retention times on the injected concentration
$C_{\mbox{\scriptsize inj}}$. The two sets of data
 correspond to different injection volumes ($V_{\mbox{\scriptsize inj}}$):
$20\,\mu$l and $100\,\mu$l. Following the procedure introduced
above for determining the average solute concentration in the
mobile phase of the migration zone $<C_i>$, we plot the
dimensionless retention parameter, $K_D$, versus $<C_i>$ in the
insert of Fig.~\ref{fig:winzor+Kd-vs-Ci}. This procedure collapses
the data from Fig.~\ref{fig:tr-vs-Cinj} onto a single curve from
which $V_{\mbox{\scriptsize inj}}$ has been removed as an
independent parameter. At the smallest concentrations in the
insert of Fig.~\ref{fig:winzor+Kd-vs-Ci}, some non-linear
dependence of $K_D$ on $<C_i>$ can be observed. We attribute this
behavior to errors introduced at the smallest signal to noise
ratios. We have not included these points in our fits.

In order to extract $B_2$ from chromatographic data, one
calculates $<C_i>$ by Eq.~\ref{eq:finalconc} and then plots $\ln
K_D$ versus $<C_i>(1-K_D)$ . Following (Nichol et al., 1978), the
slope of a linear fit to such a plot is then $2 B_2 \, N_A/M_w$,
as in Fig.~\ref{fig:winzor+Kd-vs-Ci}.

The protein concentration range typically used to measure $B_2$ by
light scattering is approximately $0<C_i<30$~mg/ml, (Velev et al.,
1998, Muschol and Rosenberger, 1995). In our SEC measurements the
protein concentrations $<C_i>$ eluting from the column, correspond
to precisely the same range, although the injected concentrations
are much higher as show in Fig.~\ref{fig:tr-vs-Cinj}. Even with
these high concentrations, we never saturated our column. Such
high injected concentrations may not be accessible for other
protein systems, and may in fact be avoided by employing larger
injection volumes, as shown by the $V_{inj}=100\mu$l data in
Fig.~\ref{fig:tr-vs-Cinj}.

\begin{figure}[ht]
\centering{\psfig{file=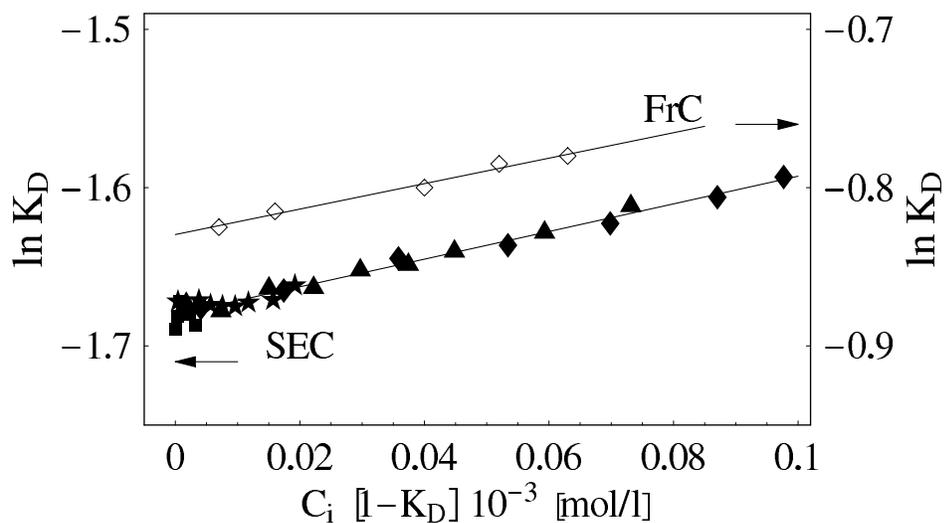,width=15cm}}
\caption{Comparison of size exclusion chromatography (SEC) and
frontal chromatography measurements for BSA. Open diamonds: BSA in
Sodium Acetate 20mM, NaCl 0.18M, pH=4.6, $A_2=1.9 \times \,
10^{-4}$ ml mol/g$^2$~(Shearwin and Winzor, 1990). Black points
are for BSA in Potassium Phosphate 50mM, pH=6.2, $A_2=2.0 \times
\, 10^{-4}$ ml mol/g$^2$. Injected concentrations are 1.14, 4.85,
10.05, 15.0, 20.7, 25.27, 30.44, 40.72, 50.99 mg/ml. Injection
volumes are for squares: $2\mu$l ; stars: $10\mu$l; triangles:
$40\mu$l; diamonds: $100\mu$l. }\label{fig:bsa-winzor+SEC}
\end{figure}

In Fig.~\ref{fig:bsa-winzor+SEC} we compare our $B_2$ results for
BSA from pulse SEC and those obtained by (Nichol et al., 1978)
using frontal chromatography. Our results show the same slope for
$\ln K_D$ as a function of $<C_i>(1-K_D)$ as those obtained by
(Shearwin and Winzor, 1990), which means the $B_2$ values are the
same. The solution conditions for the two data sets differ, but
other studies (George and Wilson, 1994, Moon et al., 2000) have
shown that $B_2$ for BSA is insensitive to many changes in
solution conditions until crystallizing conditions are approached.
Therefore we expect to measure a similar value of $B_2$. We
measured different values of $K_D$ than those in (Shearwin and
Winzor, 1990) simply because we used a different column.

\begin{figure}[ht]
\centering{\psfig{file=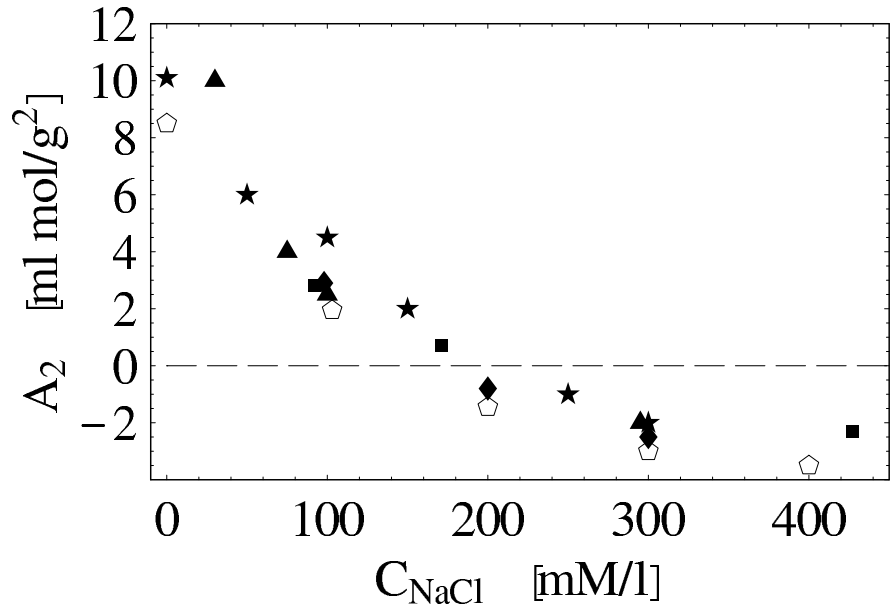,width=15cm}}
\caption{Comparison of size exclusion chromatography (SEC) and
light scattering measurements for lysozyme.
$A_2(10^{-4}$~ml~mol/g$^2$) versus NaCl concentration. Buffer:
Sodium Acetate 50~mM pH 4.7. The SEC measurements are denoted by
open pentagons. The data for 0, 100, 200, and 300 mM NaCl were
taken on a TSK column, and the data for 400~mM NaCl was with a YMC
column. The remaining data comes from published light scattering
data. Black diamonds: (Gripon et al., 1997), black triangles:
(Velev et al., 1998), black stars: (Kulkarni, 1999), black
rectangles: (Muschol et al., 1995) }\label{fig:A2-SEC-and-LS}
\end{figure}

In order to further validate the extraction of $B_2$ from size
exclusion chromatography (SEC), we compare our results to those
obtained by light scattering in Fig.~\ref{fig:A2-SEC-and-LS}, and
in Table~\ref{tab:SEC+LS}. Fig.~\ref{fig:A2-SEC-and-LS} shows the
dependence of the second osmotic virial coefficient on solution
ionic strength (added NaCl concentration) for lysozyme. Our data
agrees quantitatively with those previously obtained over a wide
range of ionic strengths. Table~\ref{tab:SEC+LS} compares $A_2$
values obtained in a different buffer, Potassium Phosphate 50mM pH
6.2. For this buffer our SEC measurements of $A_2$ also agree with
those from light scattering in their sign. The differences in
magnitude can be attributed to systematic errors associated with
light scattering and SEC measurements of $A_2$, not statistical
variation. Note that previously published results for $A_2$ from
various groups, as shown in Fig.~\ref{fig:A2-SEC-and-LS}, differ
by as much or more than the values shown in
Table~\ref{tab:SEC+LS}. These results illustrate the ability of
SEC to track the evolution of protein interactions from net
repulsive $A_2>0$ to attractive $A_2<0$.

\section*{Conclusion}

We have adapted the idea of (Nichol et al., 1978) and present
measurements of protein second virial coefficients using the
standard practice of size exclusion liquid chromatography, thereby
reducing the cost in time and material of performing $B_2$
measurements for protein solutions. After calculating the protein
concentration in the solute zone, our results agree with those
previously obtained using an independent method, light scattering,
in a number of other studies.

\section*{Acknowledgement}
Research was funded by NASA Office of Biological \& Physical
Research, Fundamental Microgravity Research in Physical Sciences
(Fluids Physics) Grant \# NAG3-2386.

\clearpage
\begin{table}[h]
\begin{center}
\begin{tabular}{c c c}
\hline
 & \hbox to 250pt {\hss $A_2\,(10^{-4}$\,ml\, mol/g$^2$)} \\
\cline{2-3}
 NaCl (mM)  & From SEC & From LS \\
\hline 0 & 2.4 & 1.8 \\
\hline 50 & 1.6 & - \\
\hline 150 & -1.0 & -1.4 \\
\hline
\end{tabular}
\end{center}
\caption{Comparison of size exclusion chromatography (SEC) and
light scattering (LS) measurements of the second virial
coefficients $(10^{-4}$~ml ~mol/g$^2$) for lysozyme in Potassium
Phosphate Buffer 50 mM, pH=6.2, at various added NaCl
concentrations.}\label{tab:SEC+LS}
\end{table}

\clearpage

\end{document}